\documentclass[final,5p,times,twocolumn]{elsarticle}

\usepackage{graphicx}
\usepackage{subcaption}

\usepackage{amsmath, amssymb}
\usepackage{amsthm}
\usepackage{hyperref}
\usepackage[export]{adjustbox}

\usepackage[mathlines]{lineno}
\newcommand*\patchAmsMathEnvironmentForLineno[1]{%
\expandafter\let\csname old#1\expandafter\endcsname\csname #1\endcsname
\expandafter\let\csname oldend#1\expandafter\endcsname\csname
end#1\endcsname
 \renewenvironment{#1}%
   {\linenomath\csname old#1\endcsname}%
   {\csname oldend#1\endcsname\endlinenomath}%
}
\newcommand*\patchBothAmsMathEnvironmentsForLineno[1]{%
  \patchAmsMathEnvironmentForLineno{#1}%
  \patchAmsMathEnvironmentForLineno{#1*}%
}
\AtBeginDocument{%
\patchBothAmsMathEnvironmentsForLineno{equation}%
\patchBothAmsMathEnvironmentsForLineno{align}%
\patchBothAmsMathEnvironmentsForLineno{flalign}%
\patchBothAmsMathEnvironmentsForLineno{alignat}%
\patchBothAmsMathEnvironmentsForLineno{gather}%
\patchBothAmsMathEnvironmentsForLineno{multline}%
}

\usepackage{xspace}
\newcommand{\uP}{u/P\xspace}
\newcommand{\vN}{v/N\xspace}
\newcommand{\emep} {$e^{+}e^{-}$}

\newcommand{\gev}{\ensuremath{\mathrm{\,Ge\kern -0.1em V}}\xspace}
\newcommand{\mev}{\ensuremath{\mathrm{\,Me\kern -0.1em V}}\xspace}
\def\mm {\ensuremath{{\rm mm}}\xspace}
\def\mum  {\ensuremath{{\,\mu\rm m}}\xspace}
\def\invfb   {\ensuremath{\mbox{\,fb}^{-1}}\xspace}
\def\invpb   {\ensuremath{\mbox{\,pb}^{-1}}\xspace}
\def\invab   {\ensuremath{\mbox{\,ab}^{-1}}\xspace}
\def\cms     {\ensuremath{{\rm \,cm}^{-2} {\rm s}^{-1}}\xspace}

\usepackage{array}
\newcommand{\PreserveBackslash}[1]{\let\temp=\\#1\let\\=\temp}
\newcolumntype{C}[1]{>{\PreserveBackslash\centering}p{#1}}
\newcolumntype{R}[1]{>{\PreserveBackslash\raggedleft}p{#1}}
\newcolumntype{L}[1]{>{\PreserveBackslash\raggedright}p{#1}}

\journal{Nuclear Instruments and Methods A}

\begin{document}

\begin{frontmatter}



\title{The Silicon Vertex Detector of the Belle II Experiment}



\author[add2211]{L.~Zani}
\ead{zani@cppm.in2p3.fr}
\author[add19]{K.~Adamczyk}
\author[add200]{L.~Aggarwal}
\author[add15]{H.~Aihara}
\author[add7]{T.~Aziz}
\author[add19]{S.~Bacher}
\author[add4]{S.~Bahinipati}
\author[add8,add9]{G.~Batignani}
\author[add222]{J.~Baudot} 
\author[add5]{P.~K.~Behera}
\author[add8,add9]{S.~Bettarini}
\author[add3]{T.~Bilka}
\author[add19]{A.~Bozek}
\author[add2]{F.~Buchsteiner}
\author[add8,add9]{G.~Casarosa}
\author[add8,add9]{L.~Corona}
\author[add12]{T.~Czank}
\author[add21]{S.~B.~Das}
\author[add222]{G.~Dujany} 
\author[add222]{C.~Finck} 
\author[add8,add9]{F.~Forti}
\author[add2]{M.~Friedl}
\author[add10,add11]{A.~Gabrielli}
\author[add10,add11]{E.~Ganiev$^{\rm B,}$}
\author[add11]{B.~Gobbo}
\author[add7]{S.~Halder}
\author[add16,add221]{K.~Hara}
\author[add7]{S.~Hazra}
\author[add12]{T.~Higuchi}
\author[add2]{C.~Irmler}
\author[add16,add221]{A.~Ishikawa}
\author[add17]{H.~B.~Jeon}
\author[add10,add11]{Y.~Jin}
\author[add19]{M.~Kaleta}
\author[add7]{A.~B.~Kaliyar}
\author[add3]{J.~Kandra}
\author[add12]{K.~H.~Kang}
\author[add19]{P.~Kapusta}
\author[add3]{P.~Kody\v{s}}
\author[add16]{T.~Kohriki}
\author[add21]{M.~Kumar}
\author[add20]{R.~Kumar}
\author[add12]{C.~La~Licata}
\author[add21]{K.~Lalwani}
\author[add2211]{R.~Leboucher}
\author[add17]{S.~C.~Lee}
\author[add5]{J.~Libby}
\author[add222]{L.~Martel} 
\author[add8,add9]{L.~Massaccesi}
\author[add7]{S.~N.~Mayekar}
\author[add7]{G.~B.~Mohanty}
\author[add12]{T.~Morii}
\author[add16,add221]{K.~R.~Nakamura}
\author[add19]{Z.~Natkaniec}
\author[add15]{Y.~Onuki}
\author[add19]{W.~Ostrowicz}
\author[add8,add9]{A.~Paladino$^{\rm A,}$}
\author[add8,add9]{E.~Paoloni}
\author[add17]{H.~Park}
\author[add2211]{L.~Polat}
\author[add7]{K.~K.~Rao}
\author[add222]{I.~Ripp-Baudot} 
\author[add8,add9]{G.~Rizzo}
\author[add7]{D.~Sahoo}
\author[add2]{C.~Schwanda}
\author[add2211]{J.~Serrano}
\author[add16]{J.~Suzuki}
\author[add16,add221]{S.~Tanaka}
\author[add15]{H.~Tanigawa}
\author[add2]{R.~Thalmeier}
\author[add7]{R.~Tiwary}
\author[add16,add221]{T.~Tsuboyama}
\author[add15]{Y.~Uematsu}
\author[add10,add11]{L.~Vitale}
\author[add15]{K.~Wan}
\author[add15]{Z.~Wang}
\author[add1]{J.~Webb}
\author[add11]{O.~Werbycka}
\author[add19]{J.~Wiechczynski}
\author[add2]{H.~Yin}
\author[]{\\ \vspace{1 mm} (Belle-II SVD Collaboration)}
\address[add1]{School of Physics, University of Melbourne, Melbourne, Victoria 3010, Australia}
\address[add2]{Institute of High Energy Physics, Austrian Academy of Sciences, 1050 Vienna, Austria}
\address[add3]{Faculty of Mathematics and Physics, Charles University, 121 16 Prague, Czech Republic}
\address[add2211]{Aix Marseille Universit$\acute{e}$ , CNRS/IN2P3, CPPM, 13288 Marseille, France}
\address[add222]{IPHC, UMR 7178, Universit$\acute{e}$ de Strasbourg, CNRS, 67037 Strasbourg, France} 
\address[add4]{Indian Institute of Technology Bhubaneswar, Satya Nagar, India}
\address[add5]{Indian Institute of Technology Madras, Chennai 600036, India}
\address[add21]{Malaviya National Institute of Technology Jaipur, Jaipur 302017, India} 
\address[add20]{Punjab Agricultural University, Ludhiana 141004, India} 
\address[add200]{Panjab University, Chandigarh 160014, India} 
\address[add7]{Tata Institute of Fundamental Research, Mumbai 400005, India}
\address[add8]{Dipartimento di Fisica, Universit\`{a} di Pisa, I-56127 Pisa, Italy,
$^A$presently at INFN Sezione di Bologna, I-40127 Bologna, Italy}
\address[add9]{INFN Sezione di Pisa, I-56127 Pisa, Italy}
\address[add10]{Dipartimento di Fisica, Universit\`{a} di Trieste, I-34127 Trieste, Italy
$^B$presently at DESY, Hamburg 22761, Germany}
\address[add11]{INFN Sezione di Trieste, I-34127 Trieste, Italy}
\address[add221]{The Graduate University for Advanced Studies (SOKENDAI), Hayama 240-0193, Japan} 
\address[add12]{Kavli Institute for the Physics and Mathematics of the Universe (WPI), University of Tokyo, Kashiwa 277-8583, Japan}
\address[add15]{Department of Physics, University of Tokyo, Tokyo 113-0033, Japan} 
\address[add16]{High Energy Accelerator Research Organization (KEK), Tsukuba 305-0801, Japan}
\address[add17]{Department of Physics, Kyungpook National University, Daegu 41566, Korea}
\address[add19]{H. Niewodniczanski Institute of Nuclear Physics, Krakow 31-342, Poland}

\begin{abstract}
Since the start of data taking in spring 2019 at the SuperKEKB collider (KEK, Japan) the Belle II Silicon Vertex Detector (SVD) has been operating reliably and with high efficiency, while providing high quality data: high signal-to-noise ratio, greater than 99\% hit efficiency, and precise spatial resolution. These attributes, combined with stability over time, result in good tracking efficiency. Currently the occupancy, dominated by beam-background hits, is quite low (about 0.5 \% in the innermost layer), causing no problems to the SVD data reconstruction. In view of the operation at higher luminosity foreseen in the next years, specific strategies aiming to preserve the tracking performance have been developed and tested on data. The time stability of the trigger allows reducing sampling of the strip-amplifier waveform. The good hit-time resolution can be exploited to further improve the robustness against the higher level of beam background. First effects of radiation damage on strip noise, sensor currents and depletion voltage have been measured: they do not have any detrimental effect on the performance of the detector. Furthermore, no damage to the SVD is observed after sudden and intense bursts of radiation due to beam losses.
\end{abstract}


\end{frontmatter}

\section*{Introduction}
\label{sec: Introduction}
The Belle II experiment~\cite{B2TechnicalDesignReport}, installed at the \emep\ SuperKEKB~\cite{SuperKEKBDesign} collider that operates mainly at the centre-of-mass energy of the $\Upsilon(4S)$ resonance, aims at probing new physics beyond the Standard Model through precise measurements on large samples of $B$ mesons, $\tau$ and charm decays. Moreover, the asymmetric energy of the 7 GeV $e^-$ beam and 4 GeV $e^+$ beam is exploited for time-dependent {\it CP}-violation measurements. SuperKEKB achieved the world’s highest instantaneous luminosity of $3.8 \times 10^{34}$ cm$^{-2}$s$^{-1}$ in December 2021 and will reach a final peak luminosity of about $6 \times 10^{35}$\cms.

Belle II already collected more than 267~\invfb and its final target is to accumulate up to 50~\invab, a 50 times larger data set than that integrated by Belle. As major upgrade of its predecessor, Belle II needs to provide similar or better performance in a harsher beam background environment. Moreover, it must achieve better vertex resolution to compensate for the reduced boost $\beta\gamma$ from 0.45 to $0.28$, and so ensure similar or better time resolution. Besides high statistics, precision measurements require precise determination of the decay vertices, excellent tracking reconstruction and energy loss measurement, as well as optimal particle identification, including for low-momentum particles. 

The Vertex Detector (VXD), as the innermost tracking device needs to cope with a hit rates of 20 (3) MHz/cm$^2$ at a radius of 14 (40) mm and requires a radiation hardness of 2 (0.2) Mrad/year. It comprises two inner layers of DEPFET pixel sensors (layer 1 and 2) that compose the Pixel Detector (PXD), and four layers (layers 3 to 6) of double-sided silicon strip detectors (DSSDs) forming the Silicon Vertex Detector (SVD)~\cite{ref:techPaper}. The remainder of this paper describes the SVD (Sec.~\ref{sec: Belle II silicon strip detector}), operational experience and performance (Sec.~\ref{sec:operation}), and the effects of beam-background and radiation (Sec.~\ref{sec:radiation}).

\section{The Belle II Silicon Vertex Detector}
\label{sec: Belle II silicon strip detector}
\begin{figure}[htbp]
     \centering
     \includegraphics[width=0.9\columnwidth]{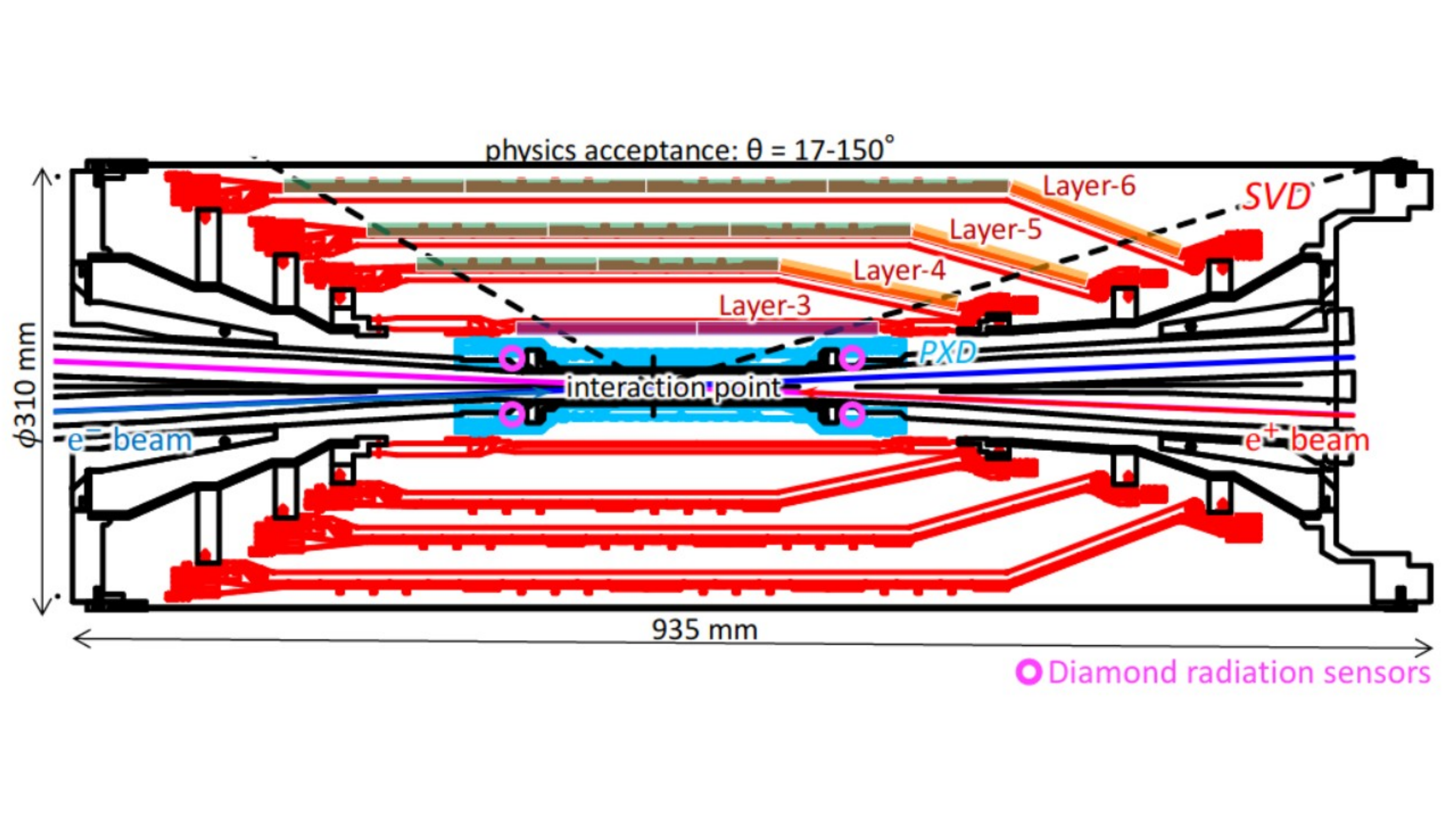}
   \caption{Schematic view of the Belle II Silicon Vertex Detector (SVD).}
 \label{fig:svdLong}
 \end{figure}

The SVD detector consists of 172 DSSD sensors arranged in four layers (3, 4, 5, and 6), which from lower to highest radius are composed of seven, ten, twelve, and sixteen ladders with two, three, four, and five sensors,
respectively. Its material budget average per layer is 0.7\% of the radiation length $X_0$. Diamond sensors are installed on the beam pipe for radiation monitoring and to trigger fast beam aborts \cite{ref:diamonds}. A longitudinal schematic view of the SVD is shown in Figure~\ref{fig:svdLong}.

Each sensor is based on an N-type bulk between 300-320 \mum thick, equipped with implanted P- and N-doped sensitive strips on opposite sides. The metal strips for the readout are AC coupled on top of the implanted strips, separated by a dielectric SiO$_2$ layer, and alternated with strips which are not readout (floating strips). Along the sensors, strips are arranged in perpendicular directions on opposite sides in order to provide 2D spatial information: the \uP side strips, orthogonal to the beam axis, measure the \(r\phi\)-direction and the \vN side provides information on the \(z\)-coordinate along the beam line.

\begin{table}[htbp]
    \centering
    \begin{tabular}{L{3.5cm}||C{0.95cm}C{0.95cm}C{1.7cm}}
        {} & Small & Large & Trap. \\
        \hline\hline
        No. \uP readout strips & 768 & 768 & 768 \\
        No. \vN readout strips & 768 & 512 & 512 \\
        Readout pitch \uP strips (\mum) & 50 & 75 & 50-75 \\
        Readout pitch \vN strips (\mum) & 160 & 240 & 240 \\
        Sensor thickness (\mum) & 320 & 320 & 300 \\
        Active Length (\mm) & 122.90 & 122.90 & 122.76 \\
        Active Width (\mm) & 38.55 & 57.72 & 57.59-38.42 \\
    \end{tabular}
    \caption{Geometrical details of the SVD DSSD sensors. All sensors have one intermediate floating strip between two readout strips.}
    \label{tab: Geometrical DSSD sensors}
\end{table}
Small rectangular sensors are used for layer 3, while layers 4, 5, and 6 are equipped with large rectangular sensors produced by Hamamatsu Photonics and slanted trapezoidal ones in the forward region produced by Micron Semiconductor. The geometrical details of the different sensors are shown in Table~\ref{tab: Geometrical DSSD sensors}. In total, the SVD covers an active area of 1.2 m$^2$ and has 224 thousands readout strips. The front-end electronic consists of APV25~\cite{ref:apv} chips that provide an analog readout of the collected signal. Each chip comprises 128 channels and consumes 0.4 W, with a fast shaping time of 50 ns and a high radiation tolerance up to 100~Mrad of integrated dose. They are operated in multi-peak mode with a 32 MHz clock, not synchronous with the bunch crossing frequency, which is about eight times higher. The signal waveform is reconstructed by six subsequent analog samplings. A mixed three/six acquisition mode has also been devised to reduce dead time, data size and occupancy at higher luminosity. While for the shorter layer 3 ladders the chips are outside the active area, for the long ladders in layers 4, 5, and 6 , the chip-on-sensor concept has been developed, to minimize the signal propagation length and to reduce the capacitance and noise. With this design, for the middle sensors of the ladders, the chips are placed only on one side of the detector, where a wrapped flex allows read out of the sensor side opposite to the chip position.
The chips are thinned to 100 \mum to minimize the material budget and stainless steel pipes for bi-phase CO$_2$ cooling at $-20^\circ$ are located on one side only. 

The SVD fulfills three main roles: it extrapolates the tracks to the PXD; it defines the region of interest for data reduction; finally, it provides standalone tracking and particle identification via the measurement of  ionisation energy loss.


\section{Operational experience and performance}
\label{sec:operation}
\begin{figure}[htp]
     \centering
     \includegraphics[width=0.9\columnwidth]{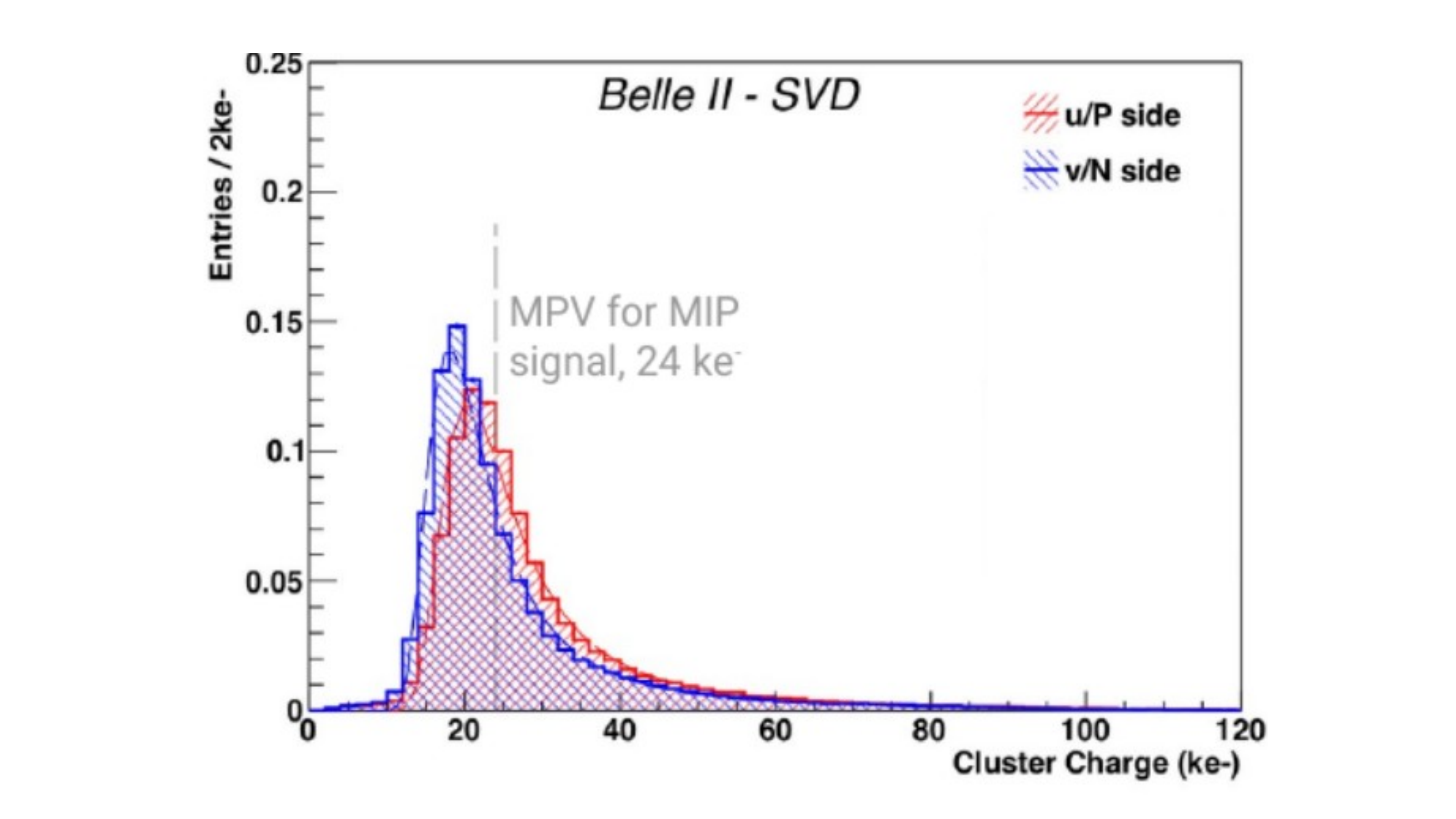}
     \includegraphics[width=0.9\columnwidth]{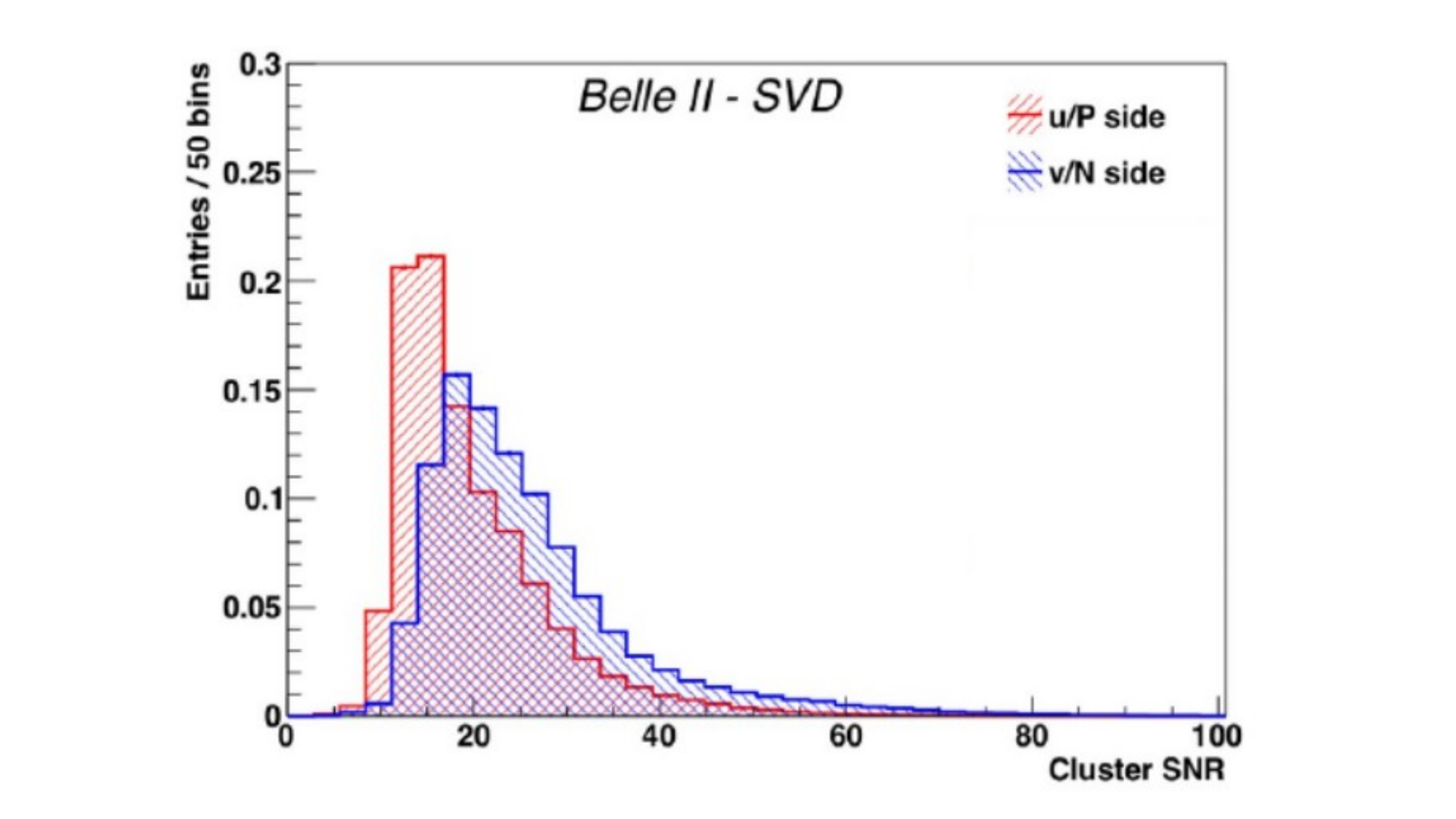}
 \caption{The cluster charge normalised with the track path length and scaled to the sensor thickness (top) and  signal-to-noise ratio (SNR) for the backward sensors of layer 3, averaged on all ladders (bottom) are shown for the u/P and v/N side in red and blue respectively. In the top plot, the dashed grey line in correspondence to 24 ke$^-$ represents the expected most probable value (MPV) for a minimal ionising particle (MIP). Data distributions are from a typical run with colliding beam in 2019 for a total luminosity of 14\invpb.}
 \label{fig:sig_charge}
 \end{figure}

The SVD has been operating smoothly with no major issues since its installation in 2019. So far, the total fraction of masked strips is less than 1\%  and only one APV25 chip out of 1748 was temporarily disabled during spring 2019, which was fully recovered after cable reconnection. Moreover temperature and calibration constants are stable, the last ones also evolving within the expected ranges, due to radiation damage. The SVD hit efficiency is measured to be above 99\% for all sensors. The signal charge normalised to the traversed sensor thickness, shown in Figure~\ref{fig:sig_charge}, is found to be in good agreement with expectations and similar in all sensors. On the u/P side the charge matches the predicted value for a minimal ionising particle when accounting for the uncertainty in the APV25 gain calibration ($\sim$15\%). On the v/N side, due to the larger pitch combined with the presence of the floating strip, approximately 10\% - 30\% signal loss is observed. 
Overall, an excellent signal-to-noise ratio (SNR) is measured in all 172 sensors with a most probable value ranging between 13 and 30. This variability in the SNR is due both to the different noise figures on the two sides of the sensors, u/P side has a larger noise due to the longer strip length and the larger inter-strip capacitance, and also to significant difference in the signal, that highly depends on the track incident angle and therefore on the sensor position. As an example, the SNR distributions for the backward sensors of layer 3, averaged over the azimuthal angle $\phi$, are shown in Figure~\ref{fig:sig_charge}. 
 
\begin{figure}[htbp]
\includegraphics[width=.65\columnwidth,keepaspectratio, right]{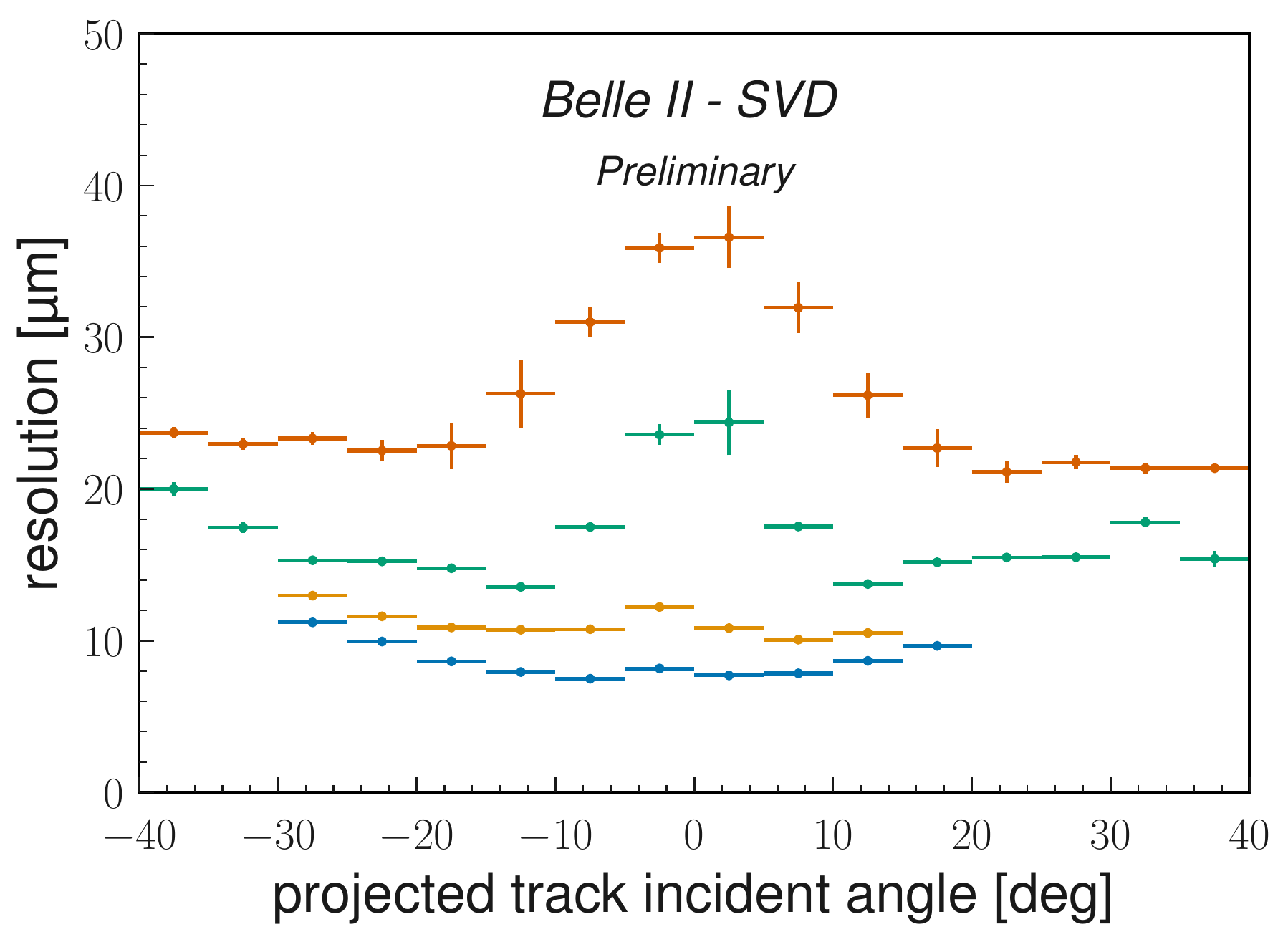}
\includegraphics[width=.98\columnwidth,keepaspectratio,right]{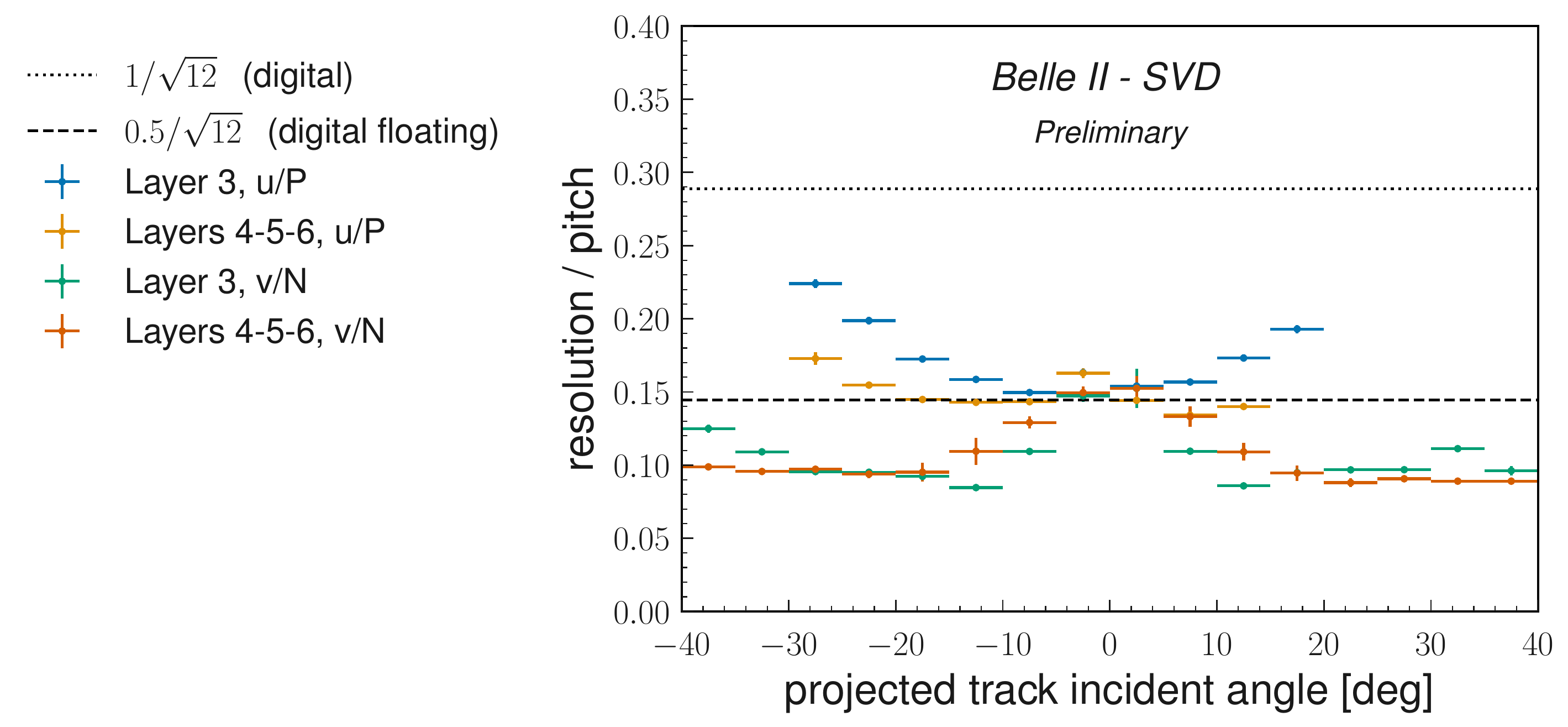}
\caption{Cluster position resolutions (top) and resolutions normalised to the pitch (bottom) as a function of the incident angle of the track traversing the sensor are shown. Results are obtained on 98\invpb collision data from 2020 run.}
\label{fig:posResol}
 \end{figure}

An excellent resolution on the cluster position is crucial to accurately reconstruct tracks and vertices. The SVD resolution is estimated from the residual of the cluster position with respect to the unbiased track extrapolation after subtracting the effect of the error on the track intercept. The measurement is performed on \emep $\rightarrow \mu^+ \mu^-$ events and results are in fair agreement with expectations from the pitch, giving 9 (11) $\mu$m for layer 3 (4, 5 and 6) u/P side and 20 (25) $\mu$m for layer 3 (4, 5 and 6) v/N side. The measured resolutions, as well as their values normalized to the pitch, are displayed as a function of the track incidence angle in Figure~\ref{fig:posResol}.

Optimal hit-time resolution of less than 2.9 (2.4) ns for u/P (v/N) side is  achieved on the SVD time, which is calibrated with respect to the event time provided by the Belle II Central Drift Chamber (CDC). The excellent time resolution can be exploited to reject off-time background hits and preliminary studies on data show a rejection power larger than 50\% on the background hits, while preserving more than 99\% signal hits. Currently, no selection based on the hit-time resolution is implemented in the reconstruction, but it will help against larger background levels expected at higher luminosities.
\section{Beam-background, occupancy limit and radiation effects}
\label{sec:radiation}
Beam-background increases the SVD hit occupancy degrading the tracking performance and causes radiation damage. A higher occupancy corresponds to a larger number of clusters that can be wrongly associated to tracks or create fake tracks in the reconstruction. The current occupancy limit, set by performance studies, is around 3\% in layer 3 and it could be
approximately doubled by exploiting the information on the hit time to reject background hits. At the present luminosity the average hit occupancy in layer 3 is below 0.5\% and it reaches 3\% at the design luminosity of $8 \times 10^{35}$ cm$^{-2}$ s$^{-1}$ according to beam-background projections based
on scaling the simulation with a data-simulation ratio. This extrapolation is affected by large uncertainties due to the assumption of optimal collimator settings and the missing contribution of the injection background not yet simulated. Moreover, the safety factor is relatively small and this motivates the proposal of a possible vertex upgrade to improve the tolerance to the hit rate and the radiation hardness. The technology assessment is currently ongoing.

The integrated dose on the SVD sensors is estimated from the correlation between the measured occupancy and the dose rate measured in the
diamond sensors \cite{ref:diamonds} and is based on several assumptions that result in an uncertainty of about 50\%. The most exposed mid plane of layer 3  collected less than 10 krad dose during the 2021 run period and 80~krad in total during the first two and a half years of data taking. It corresponds to a 1-MeV neutron equivalent fluence of $1.9\times 10^{11} $ n$_{\rm{eq}}$/cm$^2$, assuming a ratio of 2.3 $\times 10^{9}$ n$_{\rm{eq}}$/cm$^2/$krad between dose and neutron equivalent from simulation studies. 

The radiation damage has produced already observable effects, but they do not degrade the SVD performance. The increase of leakage currents are
proportional to the sensor-bulk damage due to non-ionising energy loss (NIEL). According to the NIEL model~\cite{ref:radDamage}, the damage is proportional to the 1-MeV neutron equivalent fluence, which in turn is proportional to the integrated dose. The resulting linear correlation between leakage current and integrated dose is observed for all sensors, with slopes between 2 to 5 $\mu$A/cm$^2$/Mrad, of the same order of magnitude as the effect measured by the BaBar experiment of 1 $\mu$A/cm$^2$/Mrad at 20$^{\circ}$C ~\cite{ref:babarDose}. Currently, the noise increase is dominated by strip capacitance and the leakage current contribution is highly suppressed by the short
shaping time of the APV25 chips. Also after ten years at design luminosity, corresponding to about 2 Mrad of dose with present background
extrapolation, it is not expected to significantly contribute to the total strip noise. Leakage current contribution will become comparable with the present noise after about 10 Mrad of integrated dose. 
The radiation damage is also visible in a general increase of strip noise of 20\% (30\%) on n/V (u/P) side, due to a surface effect induced by radiation on the sensors. This non-linear increase is related to fixed oxide charges in the SiO$_2$ layer increasing the inter-strip capacitance, which is expected to saturate, as it has already been observed in the v/N side. The saturation level has been reached also in the u/P side after 80 krad. No degradation of the SVD performance has been observed due to the noise increase.

Finally, non-ionising energy loss causes sensor bulk damage that can result  in a change of the effective doping and therefore of the depletion voltage. This can be monitored by measuring the v/N side noise as a function of the applied bias voltage, since N-type strips are completely insulated when the N-type bulk is fully depleted, causing also the strip noise to drop. Hence, the voltage for which the strip noise on the v/N-side reaches its minimum allows a measurement of the depletion voltage, which is shown to be constant and no changes are observed due to radiation, as expected for a low integrated 1-MeV neutron equivalent fluence accumulated so far.

\section*{Conclusions and Outlooks}
\label{sec: Conclusions and Outlooks}
The Belle II SVD has been taking data smoothly since March 2019 and shows excellent stable performance in agreement with expectations. The first effects of radiation damage have been observed at the expected level and they are not degrading the performance. The system is also ready to cope with the higher beam-background expected during future runs at higher luminosity.

\section*{Acknowledgements}
This project has received funding from the European Union's Horizon 2020 research and innovation programme under the Marie Sklodowska-Curie grant agreements No 644294 and 822070 and ERC grant agreement No 819127. This work is supported by MEXT, WPI, and JSPS (Japan); ARC (Australia); BMBWF (Austria); MSMT (Czechia); CNRS/IN2P3 (France); AIDA-2020 (Germany); DAE and DST (India); INFN (Italy); NRF and RSRI (Korea); and MNiSW (Poland).



\bibliographystyle{elsarticle-num}
\bibliography{bibliography.bib}







\end{document}